\documentclass[aps,prl,twocolumn]{revtex4}
\usepackage{graphicx}
\usepackage[T1]{fontenc} 
\usepackage{epsfig,latexsym}
\usepackage{color}
\usepackage[latin1]{inputenc}
\usepackage{amsmath}
\usepackage{amssymb}
\usepackage{amsfonts}
\usepackage{indentfirst}
\usepackage{hyperref}
\usepackage{MnSymbol,wasysym}
\usepackage{ae}
\usepackage{float}

\begin{document}
\title{Fractional Angular Momenta in Electron Beams and Hydrogen-Like Atoms}
\author{Robert Ducharme$^{1}$}
\email{robertjducharme66@gmail.com}
\author{Irismar G. da Paz$^{2}$} 
\email{irismarpaz@ufpi.edu.br}

\affiliation{$^{1}$ LVC Dynamics, 6174 Forefront Ave, Frisco, TX 75036, USA}

\affiliation{$^2$ Departamento de F\'{\i}sica, Universidade Federal
do Piau\'{\i}, Campus Ministro Petr\^{o}nio Portela, CEP 64049-550,
Teresina, PI, Brazil}

\begin{abstract}
In an earlier letter [Ducharme \textit{et al.} Phys. Rev. Lett. \textbf{126}, 134803 (2021)], a solution to the Dirac equation for a relativistic Gaussian electron beam showed that for a diverging beam the spin of each electron is the sum of fractional contributions from both the spin angular momentum (SAM) and orbital angular momentum (OAM) operators. Fractional angular momenta emerge when  eigenstates of the Dirac equation can be decomposed into two terms of opposite spin. Each of these terms being eigenstates of both the SAM and OAM operators. Building on this understanding, the same method used to calculate fractional angular momenta in beams is applied here to solutions of the Dirac equation for hydrogen-like atoms. The results strengthen the idea that factorization of the Klein-Gordon equation using Dirac matrices equation does more than introduce spin, it also produces a specific mixing of the angular momentum states leading to the presence of fractional angular momenta and related effects such as fractional Gouy phase in the case of beams and fractional wave-particle duality in both atoms and beams.
\end{abstract}

\maketitle

{\textit{Introduction}.---} 
In 1992, a team led by Leslie Allen made the discovery of orbital angular momentum (OAM) in light beams \cite{review-twisted-light} meaning that the total angular momentum (TAM) in light is the sum of spin angular momentum (SAM) of individual photons and intrinsic \cite{Bliokh:23} OAM that twists around the axis of the beam. The discovery of OAM in light further led to its prediction in electron beams \cite{Bliokh:07} and its subsequent experimental realization \cite{EVB-exp}. The potential to exploit OAM in multiple diverse fields including microscopy, communications, radar and particle manipulation is driving many current research efforts. There is nevertheless an interesting theoretical result that has emerged in the case of tightly-focused (non-paraxial) beams that the expectation value the OAM operator is not an integer multiple of the reduced Planck constant $\hbar$. Similarly, the magnitude of the expectation value SAM operator can be less than $\frac{\hbar}{2}$. This has led to the emergence of the terms fractional SAM (FSAM) and fractional OAM (FOAM) in the quantum optics literature. For example, Bloikh \textit{et. al.} have studied Bessel beams leading to exact non-paraxial solutions to both Maxwell's equations for light beams \cite{Bliokh:10} and the Dirac equation for electron beams \cite{Bliokh:11} that provide explicit formula for FSAM and FOAM and a clear understanding in terms of Berry phase \cite{Berry}. As theoretical interest in the interplay of SAM and OAM continues \cite{RD21}, there is growing number of experimental approaches to measuring and exploiting it \cite{FOAM-exp}. From the quantum field theorist perspective, Fukushima and Pu \cite{Fukushima21} have expressed surprise at results from electron beam studies \cite{SMB} and acknowledged the need for more research into the issues raised. 

Here, the objective is to move beyond beams and investigate whether FOAM and FSAM exist in hydrogen-like atoms. Hydrogen-like atoms have a variable atomic number $Z$ but only one electron. Electrons in the orbitals of atoms like those in beams must satisfy the Dirac equation. FOAM and FSAM emerge when eigenstates of the Dirac equation decompose into two terms of opposite spin. Each of the two terms being an eigenstate of both the SAM and OAM operators. It is clear, therefore, that the factorization of the Klein-Gordon equation using Dirac matrices does more than introduce spin, it can also generates a mixing of two angular momentum states. In electron beams, FOAM and FSAM have related effects such as fractional Gouy and Berry phases \cite{RD21}. It will be argued next that FOAM and FSAM more generally lead to fractional wave-particle behavior.

In standard Compton scattering, it is usual to represent the electron as a point-like particle since the outcome does not depend on the size of the electron wave packet \cite{Corson11}. If, however, integer OAM is present in the wave packet, angular momentum conservation requires that the electron must be treated as a wave \cite{Liao25}. Thus, if the electron has FOAM but no integer OAM, then it reasonable to hypothesize the electron transitions from being particle-like as FOAM tends to zero to becoming wave-like as FOAM tends to $\hbar$. Clearly, it follows from this argument that particle-to-wave transition is controllable through the device that encodes FOAM into the electron wave packet.

Greenberger and Yasin \cite{Greenberger88} have quantified wave-particle duality for path distinguishability in interferometry using the information-theoretic inequality $\mathcal{P}_{GY}+\mathcal{W}_{GY}\le 1$ where $\mathcal{W}_{GY}$ denotes wave-ness and $\mathcal{P}_{GY}$ is particle-ness for beam splitting experiments. This relation builds on an earlier formulation of the Bohr complimentarity principle \cite{Bohr28} by Wootters and Zurek \cite{Wootters79} and has been shown \cite{Jaeger95} to simplify to $\mathcal{P}_{GY}+\mathcal{W}_{GY} = 1$ for pure single particle states. Wave-particle duality has also been extensively studied \cite{Friedman88, Gover18} across the rich spontaneous and stimulated radiative emission field. The consensus finding between both interferometry and radiative emissions studies \cite{Pan18} being that the particle-to-wave transition is continuous and controllable through the devices that shape the electron wave packet. Our argument that FOAM in electron beams also facilitates controllable particle-to-wave transition therefore adds to this growing consensus, if substantiated.

As stated earlier, FOAM and FSAM are just the expectation values of the respective OAM and SAM operators. Thus, to pass from a calculation of fractional angular momenta in beams to one in a hydrogen-like atom is all that is needed is to replace the beam function for a hydrogen-like atom wave function.  Although beam function solutions to the Dirac equation are relatively recent advances, solutions to the Dirac equation for hydrogen-like atoms have been available since 1928 \cite{Darwin28}. It will be shown through this approach that significant FOAM is present in the innermost shells of atoms that have large values of Z. For comparison, to maximize FOAM in electron beams a stream of energetic electrons must be tightly focused to decrease the size of the beam waist and increase its opening angle. It is therefore tight localization of fast electron in both atoms and beams that leads to the most FOAM. Finally, it will be demonstrated that like beams pure eigenstates of the Dirac equation for the hydrogen-like atom each decompose into two orthogonal eigenstates of the SAM and OAM operators that have opposite spins to each other. The probabilities of the electron existing in each of these states is calculated and shown to be related to FOAM.

{\textit{FOAM in Hydrogen-Like Atoms}.---} 
Consider a hydrogen-like atom containing an electron of charge $-e$, reduced mass $\mu$, 3-momentum $\mathbf{p}$ and total energy $E$ bound to an electrostatic potential of charge $Ze$. Let $\mathbf{x}$ be the 3-position of the atomic nucleus at time $t$. Further, let $\mathbf{r}$ and $\tau$ denote the spatial and temporal displacements of the electron from the nucleus. It is usual in solving the Dirac equation for this problem to assume $\mathbf{x} = \mathbf{p} = \tau = 0$. The Dirac bispinor for the electron can then be written in terms of spherical polar coordinates \cite{Bjorken65} as
\begin{eqnarray} 
\label{eq: general_solution}
\Psi_{n, \kappa, m_j}(\mathbf{r}, t) = 
\frac{1}{r}\left[ \begin{array}{c}
g_{n,\kappa}(r) \Omega_{\kappa, m_j}(\theta, \phi)
\\
\imath f_{n, \kappa}(r) \Omega_{-\kappa, m_j}(\theta, \phi)
\end{array} \right] e^{-\imath E t}
\end{eqnarray}
where $f_{n\kappa}(r)$ and $g_{n\kappa}(r)$ are functions of the radial coordinate $r$,
\begin{multline} 
\label{eq: weyl_spinor}
\Omega_{\kappa,m_j}(\theta, \phi) = 
\sqrt{\frac{\kappa + \frac{1}{2}-m_j}{2\kappa+1}} Y_{\kappa, m_j-1/2}(\theta, \phi)\chi_+
\\
-\text{sgn} (\kappa)\sqrt{\frac{\kappa + \frac{1}{2}+m_j}{2\kappa+1}}  Y_{\kappa, m_j+1/2}(\theta, \phi)\chi_-
\end{multline}
is the Weyl spinor, $\theta$ is the polar angle and $\phi$ is the azimuth angle, $Y_{l, m_j}$ is the spherical harmonic function, $\chi_+ = (1 \, 0)^T$ and, $\chi_- = (0 \, 1)^T$ are spinors and $T$ is transpose. There are seven different quantum numbers applicable to central solutions of the Dirac equation in common usage. These include the principal quantum number $n = 1,2,3, ...$, the azimuth quantum number $l = 0, 1, ... ,  n-1$, the spin quantum number $s=\frac{1}{2}$, the two-valued spin quantum number $m_s=\pm\frac{1}{2}$, the total angular momentum quantum number $j$ that takes all half-integer values on the range $|l-s| \le j \le l+s$, the secondary total angular momentum quantum number $m_j$ that takes all integer values on the range $-j \le m_j \le j$, and the integer  
\[
  \kappa = \begin{cases}
    -l-1 & \text{if $j = l+1/2$} \\
    \quad l & \text{if $j = l-1/2$}
  \end{cases}
\]
To determine the dependence of fractional angular momenta on Z and n, it is necessary to hold $\kappa$ and $m$ constant. For this purpose, it will be convenient to set $\kappa = -n$ and $m_j = \pm(n-\frac{1}{2})$. In particular, the $\kappa = -n$ setting leads to simpler expressions for the un-normalized radial functions  $f_{n,\kappa}(r)$ and $g_{n,\kappa}(r)$ than other values of $\kappa$. These are
\begin{eqnarray}
\label{eq: radial_func_f}
f_{n,-n}(r) = Z \alpha r^\eta_n e^{-k_n r}
\\
\label{eq: radial_func_g}
g_{n,-n}(r) = (n+\eta_n)r^\eta_n e^{-k_n r}
\end{eqnarray}
where $k_n = \frac{Z\alpha \mu c}{n\hbar}$ is the radial decay constant, $c$ is the velocity of light, $\alpha$ is the fine structure constant and $\eta_n = \sqrt{n^2-Z^2\alpha^2}$. Putting Eqs. (\ref{eq: general_solution}) through (\ref{eq: radial_func_g}) together then gives
\begin{equation}
\label{eq: constrained_solution}
\Psi_{n,\pm}(\mathbf{r}, t) = 
 \left[ \begin{array}{c}
(n+\eta_n)\chi_\pm
\\
\pm \imath Z\alpha (\cos \theta \chi_\pm + \sin \theta e^{\pm \imath \phi} \chi_\mp )
\end{array} \right]\psi_{n, \pm}(r,t)
\end{equation}
where $\Psi_{n,\pm}(\mathbf{r}, t)=\Psi_{n, -n, \pm n \mp 1/2}(\mathbf{r}, t)$,
\begin{equation}
\psi_{n,\pm}(r,t) = \mathcal{N}_n \rho^{\eta_n-1} (\sin \theta e^{\pm \imath \phi})^{n-1} e^{ - \frac{\rho}{2}-\frac{\imath}{\hbar} E t}
\end{equation}
$\mathcal{N}_n$ is the normalizing constant and $\rho = 2k_nr$ is a dimensionless form of the radial coordinate. This expression represents a family of bispinor solutions to the Dirac equation describing the $1S_{1/2}$, $2P_{3/2}$, $3D_{5/2}$ and  $4F_{7/2}$ electron orbitals as well as all the higher level orbitals in the same series. The energy eigenvalues for these solutions are $E_n = \mu c^2\frac{\eta_n}{n}$. 

The axial component of the SAM operator is $\hat{S}_3 = \text{diag}(\sigma_3, \sigma_3)$ where $\sigma_3$ is a Pauli matrix. The expectation value of $\hat{S}_3$ for  wave function $\Psi_{n\pm}$ can thus be written
\begin{equation}
\langle \Psi_{n,\pm}^\dagger \hat{S}_3 \Psi_{n,\pm} \rangle =
2 \pi \int_0^{\pi} \!\!\! \int_0^\infty  \Psi_{n,\pm}^\dagger \hat{S}_3 \Psi_{n,\pm} r^2 dr \sin\theta d \theta 
\end{equation}
having chosen $\langle \Psi_{n,\pm}^\dagger \Psi_{n,\pm} \rangle = 1$ to be the normalizing condition. Thus, FSAM, OAM, FOAM and total angular momentum (TAM) can be determined from the expressions
\begin{eqnarray}
FSAM = \langle \Psi^\dagger \hat{S}_3 \Psi \rangle,
\,\,
TAM = \langle \Psi^\dagger (\hat{S}_3+\hat{L}_3) \Psi \rangle,
\\
OAM =  \lfloor \langle \Psi^\dagger \hat{L}_3 \Psi \rangle \rfloor,
\,\,
FOAM = \langle \Psi^\dagger \hat{L}_3 \Psi \rangle - OAM,
\end{eqnarray}
where $\hat{L}_3 = \hbar \frac{\partial}{\partial \phi}$ is the axial component of the OAM operator and the floor function $\lfloor x \rfloor $ returns the integer part of $x$. Evaluating each of these integrals for the hydrogen-like wave function $\Psi = \Psi_{n, \pm}$ then reveals
\begin{eqnarray}
FOAM_A(Z,n,s) = \frac{Z^2\alpha^2}{(n+\eta_n)^2+Z^2\alpha^2} 
\frac{4nm_s\hbar}{2n+1},
\\
FSAM_A(Z,n,s) = \frac{(n+\eta_n)^2 - \frac{Z^2\alpha^2(2n-1)}{2n+1}}{(n+\eta_n)^2+Z^2\alpha^2} 
m_s\hbar
\end{eqnarray}
alongside the two better known results $OAM_A = \pm(n-1) \hbar = l \hbar$ and $TAM_A = \pm(n-1/2) \hbar$. Note, the subscript $A$ signifies the expectation values have been calculated for the hydrogen-like atom to distinguish the results from those for electron beams that will be presented in the next section. 

{\textit{Origin of FOAM in Atoms and Beams}.---}
Fractional angular momenta in electron beams have been calculated in previous work for both Bessel \cite{Bliokh:11} and Gaussian beams \cite{RD21} giving
\begin{eqnarray}
FOAM_B(\theta_D,E_B,s) = \left( 1-\frac{mc^2}{E_B}\right) \sin^2 \theta_D m_s\hbar,
\\
FSAM_B(\theta_D,E_B,s) = \left[ 1 - \left( 1-\frac{mc^2}{E_B}\right) \sin^2 \theta_D \right]l
\end{eqnarray}
where $m$ is the rest mass of an electron having total energy $E_B$ and $\theta_D$ is the opening angle of the beam. Figure 1 shows a comparison of FOAM and FSAM in a hydrogen-like atom (top) and an electron beam (bottom). It can be seen that FOAM  increases with the atomic number $Z$ in atoms as it does with the opening angle $\theta_D$. The connection between $Z$ and $\theta_D$ being that high values of either indicate tight spatial localization. Specifically, electron orbitals are smaller in the presence of a highly charged nucleus whereas the strong focusing of an electron beam that produces a large opening angle also results in a small beam waist. Additionally, it is clear from the graphs that FOAM depends on the energy of the electron through the principal quantum number $n$ in atoms and $E_B$ in beams. In summary, it can therefore be said that FOAM is largest in the presence of energetic electrons that have become tightly localized in at least two space dimensions for both atoms and beams.

\begin{figure}[htp]
	\centering
	\includegraphics[width=8.2 cm]{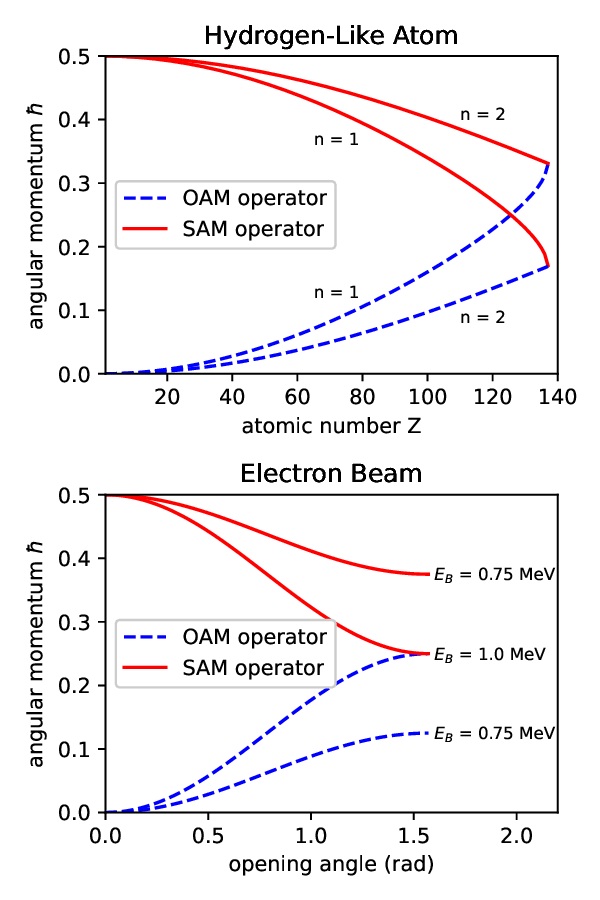}
	\caption{Expectation values for the fractional spin and orbital angular momenta of electron in a hydrogen-like atom (top) and electron beam (bottom).}
	\label{fig:1}
\end{figure}

Insight into the mathematical origin of FOAM can be obtained from the ground state solution $\Psi_0\pm$ of the Dirac equation for the hydrogen-like atom. This solution has no integer OAM. So, observe from Eq. (\ref{eq: constrained_solution}) for the $n=1$ case, that only one of the four components in the bispinor contributes to FOAM. Figure 2 shows the twisted wave front for this component within a radius $R$ of the atom such that the phase angle $-\frac{1}{\hbar}Et \pm \phi$ for each point on the wave front has the same value. For comparison, the other three components of $\Psi_\pm$ have planar wave fronts. For $n>1$, Eq. (\ref{eq: constrained_solution}) shows that integer OAM comes from the scalar part of the wave function meaning that all of the bispinor components have twisted wave fronts. In this case, FOAM emerges when one of the bispinor components happens to be more or less twisted than the others. This is one perspective but there is a second one. In the treatment of electron beams \cite{Bliokh:11, RD21} it is usual to think of the magnetic lens has a creating a mixed beam state where different eigenstates of both the SAM and OAM operator sum to give the beam function. We will therefore next develop the idea that the wave functions of atomic orbitals can be similarly decomposed into eigenstates of the SAM and OAM operators.

\begin{figure}[htp]
	\centering
	\includegraphics[width=8.2 cm]{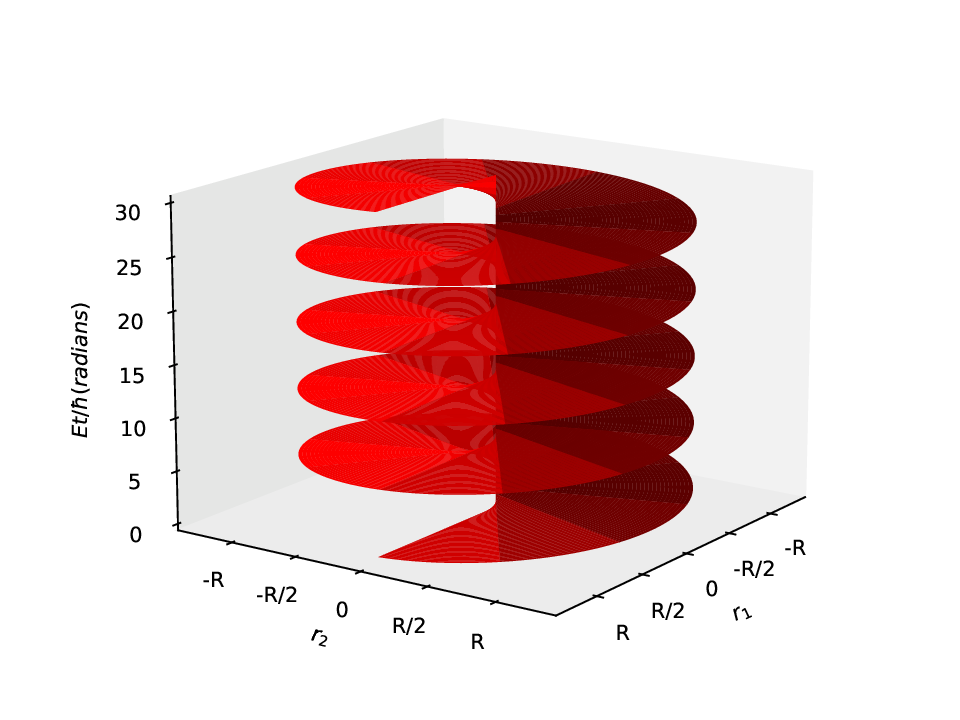}
	\caption{Twisted wavefront for one bi-spinor component in a hydrogen-like atom wave function carrying fractional OAM. The other three components of the bi-spinor will have planar wavefronts unless the electron also has integer OAM}
	\label{fig:2}
\end{figure}
From inspection, Eq. (\ref{eq: constrained_solution}) is readily separated into two terms that have opposite spins. This gives $\Psi_{n,\pm} = \Psi_{n,\pm}^A + \Psi_{n,\pm}^B$ where
\begin{eqnarray}
\label{eq: eigenstate_A}
\Psi_{n,\pm}^A =  \left[ \begin{array}{c}
n+\eta_n
\\
\pm \imath Z\alpha \cos \theta 
\end{array} \right]\psi_{n,\pm}\chi_\pm,
\\
\label{eq: eigenstate_B}
\Psi_{n, \pm}^B =  \left[ \begin{array}{c}
0
\\
\pm \imath Z\alpha \sin \theta e^{\pm \imath \phi} 
\end{array} \right]\psi_{n,\pm} \chi_\mp
\end{eqnarray}
It can then be shown that
$\hat{S}_3 \Psi_{n,\pm}^A = \pm \frac{\hbar}{2}\Psi_{n,\pm}^A$, $\hat{S}_3 \Psi_{n,\pm}^B = \mp \frac{\hbar}{2}\Psi_{n\pm}^B$, $\hat{L}_3 \Psi_{n,\pm}^A = \pm(n-1) \hbar \Psi_{n,\pm}^A$ and $\hat{L}_3 \Psi_{n,\pm}^B = \pm n\hbar\Psi_{n,\pm}^B$ confirming that $\Psi_{n, \pm}^A$ and $\Psi_{n,\pm}^B$ are indeed eigenstates of both the SAM and OAM operators. Further, it can be seen $\Psi_{n, \pm}^A$ and $\Psi_{n,\pm}^B$ are orthogonal to each other based on the identity $\chi{\pm}^T\chi{\mp}=0$. These results mean that an electron in an orbital $\Psi_{n, \pm}$ has a probability 
\begin{equation}
\label{eq: prob_dens_A}
\langle \Psi_{n,\pm}^{A \dagger}\Psi_{n,\pm}^A \rangle = \frac{(n+\eta_n)^2+  \frac{Z^2\alpha^2}{(2n+1)}}{(n+\eta_n)^2+Z^2\alpha^2}  = 1-\frac{|FOAM_A|}{\hbar}
\end{equation}
of being in the eigenstate $\Psi_{n,\pm}^A$ with $OAM=\pm(n-1) \hbar$ and $SAM=\pm\frac{\hbar}{2}$ having used Eqs. (\ref{eq: eigenstate_A}) and (\ref{eq: eigenstate_B}). Similarly, the probability of the same electron being in the eigenstate $\Psi_{n,\pm}^B$ with $OAM=\pm n \hbar$ and $SAM=\mp\frac{\hbar}{2}$ is
\begin{equation}
\label{eq: prob_dens_B}
\langle \Psi_{n,\pm}^{B \dagger}\Psi_{n,\pm}^B \rangle = \frac{Z^2\alpha^2}{(n+\eta_n)^2+Z^2\alpha^2} 
\frac{2n}{(2n+1)} = \frac{|FOAM_A|}{\hbar}
\end{equation}
Eqs. (\ref{eq: prob_dens_A}) and (\ref{eq: prob_dens_B}) therefore show that it is the probability of an electron in an orbital being in the angular momentum eigenstate $\Psi_{n,\pm}^B$ that determines both the values of FSAM and FOAM.

Figure 3 depicts the radial probabilities for the $\Psi_{1, +}^A$ and $\Psi_{1, +}^B$ eigenstates as a function of position. The sum of these eigenstates being the Dirac wave function $\Psi_{1,+}$ that describe a spin-up electron in the $1S_{1/2}$ orbital of a lead atom (Z=82). The lead atom has been chosen since the probability of electron being in the $\Psi_{1, +}^B$ eigenstate is vanishingly small for much lighter nuclei. It is interesting that the electron in the $\Psi_{1, +}^B$ eigenstate has OAM and must therefore be assumed to be wave-like in interactions with high energy photons. The probability of this happening is $\mathcal{W}_{AM} = \frac{FOAM}{\hbar}$. Conversely, the probability of the electron being in the particle-like $\Psi_{1, +}^A$ eigenstate is just $\mathcal{P}_{AM} = 1-\mathcal{W}_{AM}$. 

\begin{figure}[htp]
\centering
\includegraphics[width=8.2 cm]{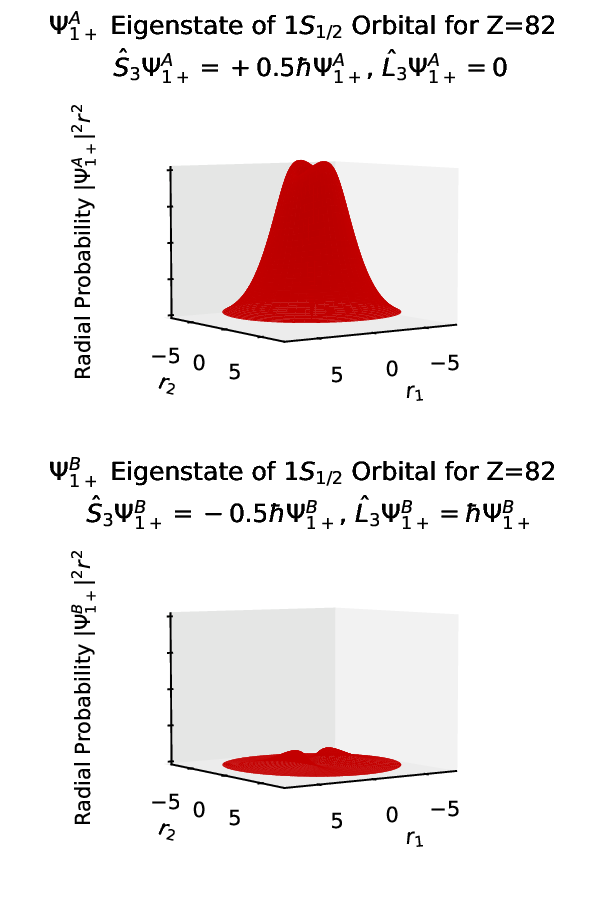}
\caption{Radial probabilities for the $\Psi_{1+}^A$ (top) and $\Psi_{1+}^B$ (bottom) eigenstates of the SAM and OAM operators for a spin-up electron in the $1S_{1/2}$ orbital of a lead atom.}
\label{fig:3}
\end{figure}

{\textit{Concluding Remarks}.---}
The total angular momentum of an electron in a paraxial beam comprises the half-integral spin of the electron particle plus an integer contribution from OAM. If, however, the beam is tightly focused using a magnetic lens the half-integral component of the angular momentum fractionates into FSAM and FOAM that emerge as the expectation values of the SAM and OAM operators. The implication being that the electron is distributed between two possible eigenstates of the SAM and OAM operators that have opposite spins. If one of these states has zero integer OAM, there is a probability $\mathcal{P}_{AM}$ that the electron may behave like a particle. By contrast, there is also a complimentary probability $\mathcal{W}_{AM} = 1-\mathcal{P}_{AM}$ that the electron will have OAM, thus, requiring the electron to be treated as a wave. We have inferred from this that OAM-related wave-particle duality in electron beams fits a signature pattern also observed in interferometry and radiative emission studies that the particle-to-wave transition is controllable through methods and devices used to shape the electron wave packet. In this case we see that the FOAM related controllable transition can be explained using the Dirac equation but not the Schrodinger or Klein-Gordon equations. 

FSAM and FOAM has been calculated for hydrogen-like atoms using the same method that works for electron beams. This reveals that FOAM becomes significant in atoms in the innermost orbital of a highly charged nuclei but there is almost no FOAM in light nuclei. It follows that FOAM emerges in both atoms and beams when fast electrons are tightly localized in at least two space dimensions. In an atom, FOAM is controllable through the atomic number of the nucleus and its quantum state. We have shown the solutions of the Dirac equation that describe orbitals can be expressed in terms of two eigenstates of the SAM and OAM operators similar to electron beams. We have further shown that the probability of the electron existing in either one of these states determines the values of both FOAM and FSAM. Thus, our reasoning further predicts, for example, that standard Compton scattering off the ground state shell of a hydrogen-like atom would be a better approximation for a light nucleus than a heavy one owing to the presence of more FOAM in the heavy atom. Overall, we conclude eigenstates of the Dirac equation can be expressed in terms of an alternative angular momentum basis set but more work will be needed to fully investigate its properties. 
\begin{acknowledgements}
I. G. da Paz thanks Grant No. 306528/2023-1 from CNPq.
\end{acknowledgements}

\end{document}